\documentclass[10pt]{llncs}
\usepackage[colour, compact]{eventB}

\usepackage[colour]{btheory}
\usepackage{xcolor}

\usepackage[utf8x]{inputenc}
\definecolor{ao(english)}{rgb}{0.0, 0.5, 0.0}
\DeclareUnicodeCharacter{8473}{{\(\mathbb{R}\)}}

\usepackage{subfig}
\usepackage{standalone}
\usepackage{url}
\usepackage{amssymb}
\usepackage{graphics}

\usepackage{amsmath}

\usepackage{bsymb}

\usepackage{spacing}

\usepackage{alltt}
\usepackage{framed}

\usepackage{marginnote}
\usepackage{tikz}

\title{Theory Plug-in for Rodin 3.x}
\author{T.S.~Hoang\inst{1} \and L.~Voisin\inst{2} \and A.~Salehi\inst{1} \and M.~Butler\inst{1} \and T.~Wilkinson\inst{1} \and N.~Beauger\inst{2}}
\institute{
  ECS, University of Southamtpon, U.K.
  \and
  Systerel, France
}

\begin{document}
\maketitle

\begin{abstract}
  The Theory plug-in enables modellers to extend the mathematical modelling notation for Event-B, with accompanying support for reasoning about the extended language. Previous version of the Theory plug-in has been implemented based on \emph{Rodin 2.x}.  This presentation outline the main improvements to the Theory plug-in, to be compatible with Rodin \emph{3.x}, in terms of both reliability and usability.  We will also present the changes that were needed in the Rodin core to accommodate the Theory plug-in.  Finally, we identify future enhancements and research directions for the Theory plug-in.
\end{abstract}

\section{Introduction}
\label{sec:introduction}

The Theory plug-in \emph{v3.0} has been implemented based on \emph{Rodin 2.x}. The Theory plug-in enables modellers to extend the mathematical modelling notation for Event-B, with accompanying support for reasoning about the extended language.  As a result for this extension, different formulae within an Event-B model can be built from different formula factories. As an example, when dealing with a proof obligation, in particular, checking if an existing proof is reusable, we have to deal with two formula factories: (1) the factory associated with the proof obligation (coming from the model), and (2) the factory associated with the existing proof.  In the case where the model is altered, these formula factories might be incompatible, rendering the existing proof unusable.  In Rodin 3.0, there are major changes (compared to Rodin 2.8) within the Rodin Core~(\url{http://wiki.event-b.org/index.php/Rodin_Platform_3.0_Release_Notes}).

\begin{itemize}
\item \textbf{Stronger AST Library}: The API of the AST library has been strengthened to mitigate risks of unsoundness when mixing several formula factories. Now, every AST node carries the formula factory with which it was built, and the AST operations that combine several formulas check that formula factories are compatible.

\item \textbf{Stronger sequent prover}: In order to improve the reliability of the proof status when working with mathematical extensions, the reasoners can be declared as context-dependent. The proofs that use a context dependent reasoner will not be trusted merely based on their dependencies, but instead they will be replayed in order to update their status. This applies in particular to Theory Plug-in reasoners, that depend on the mathematical language and proof rules defined in theories, which change over time.
\end{itemize}
The above changes directly effect the Theory plug-in, in particular to avoid building formulae from sub-formulae with different formula factories.

\section{Mathematical Extensions via the Theory Plug-in} 
\label{sec:background}
The Theory plug-in~\cite{DBLP:conf/birthday/ButlerM13} enables developers to define new (polymorphic) data types and operators upon those data types.  These additional modelling concepts might be defined axiomatically or directly (including inductive definitions).  We have made use of the Theory plug-in capability to define domain-specific concepts and provide proof rules for reasoning about them.  We focus on the following features of the Theory plug-in that are relevant for our report.

\subsection{Modelling extensions} The following modelling extensions are offered by the Theory plug-in: \emph{Polymorphic Inductive Datatypes}, \emph{Axiomatic Datatypes}, \emph{Operators (using Direct Definition, Inductive Definition or Axiomatic definition)}

\subsubsection{Polymorphic Inductive Datatypes} 
\begin{figure}[!h]
    \centering
    \includegraphics[scale=0.4]{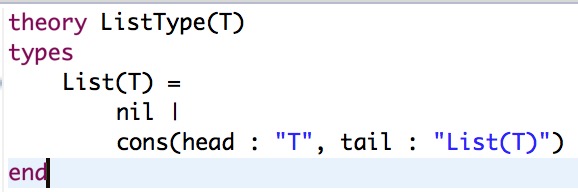}
    \caption{\texttt{List} Datatype}
     \label{fig:list-datatype}
  \end{figure}

A datatype with type parameters (polymorphic) is defined using several constructors. Each constructor can have zero or more destructors.  As an example, the common \textit{\textcolor{ao(english)}{List}} inductive datatype is defined as in \Fig{fig:list-datatype}. A List is either an empty list ({\textcolor{ao(english)}{nil}}) or a concatenation of an element ({\textcolor{ao(english)}{head}}) and a list ({\textcolor{ao(english)}{tail}}). 

\subsubsection{Axiomatic Datatypes} 
\begin{figure}[!h]
    \centering
    \includegraphics[scale=0.4]{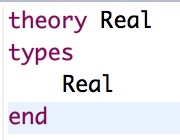}
    \caption{\texttt{Real} Datatype}
    \label{fig:real-datatype}
  \end{figure}

A datatype without any definition is axiomatically defined.  By convention, an axiomatic datatype satisfies the \emph{non-emptiness} and \emph{maximality} properties, i.e., for an axiomatic type $S$,
  \begin{equation}
    \label{eq:non-emptiness}
    S \neq \emptyset \tag{non-emptiness} \\
  \end{equation}
  \begin{equation}
    \label{eq:maximality}
    \forall x \qdot x \in S~. \tag{maximality} \\
  \end{equation}
An example of a axiomatic type for Real number (without any additional axioms) is in \Fig{fig:real-datatype}.

\subsubsection{Operators} 
\begin{figure}[!h]
   \centering
   \includegraphics[scale=0.4]{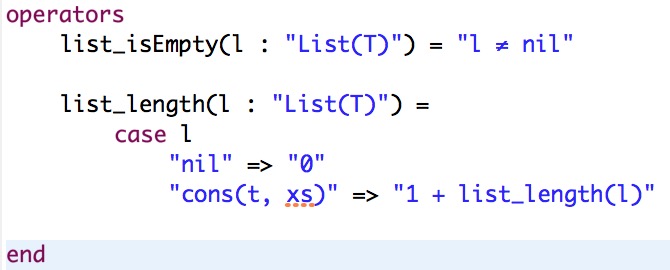}
   \caption{Some operators for \texttt{List} Datatype}
   \label{fig:list-ops}
 \end{figure}

\begin{figure}[!h]
   \centering
   \includegraphics[scale=0.4]{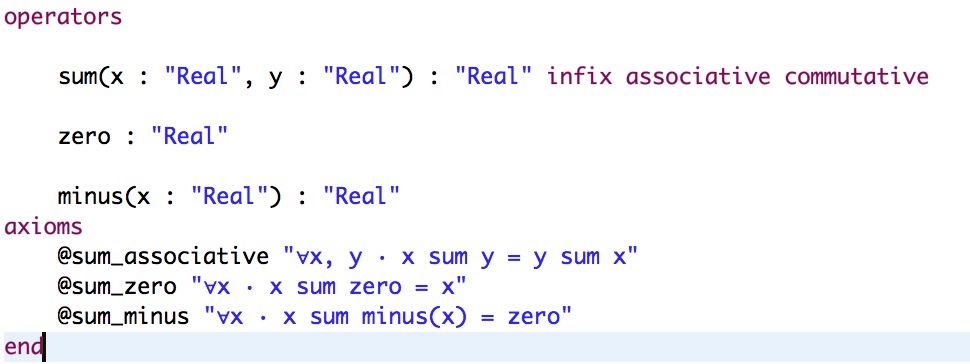}
   \caption{Some operators for \texttt{Real} Datatype}
   \label{fig:real-ops}
 \end{figure}

Operator can be defined \emph{directly}, \emph{inductively} or \emph{axiomatically}.  \Fig{fig:list-ops} shows the definitions of two operators \texttt{list\_isEmpty} (defined directly) and \texttt{list\_length} (defined inductively).
An operator defined without any definition will be defined axiomatically. Operator notation can be either \texttt{PREFIX} (default) or \texttt{INFIX} (for operator with two or more arguments). Further properties can be declared for operators include \emph{associativity} and \emph{commutativity}. \Fig{fig:real-ops} shows the declaration for three operators: \texttt{sum}, \texttt{zero}, and (unary) \texttt{minus}. In particular, \texttt{sum} is declared to be an infix operator which is associative and commutative.  The axioms are the assumption about these operators that can be used to defined proof-rules.

\subsubsection{Reasoning extensions}
To support reasoning about the user-defined datatypes and operators, the Theory plug-in offers 5 different \emph{proof tactics}:
\begin{itemize}
\item Manual inferencing
\item Manual rewriting (including expanding definitions)
\item Automatic inferencing
\item Automatic rewriting
\item Automatic definition expanding
\end{itemize}
These proof tactics are configured using the proof rules as a part of the user-defined theories. The manual tactics corresponding to application of a single proof rule or expanding a single definition.  For rewriting tactics, they often work at a particular location of the formula.  Automatic applications of  proof rules mean to repeatedly apply one or more rules until no progress can be made.  There are two type of proof rules supported by the Theory plug-in, namely, rewrite rules and inference rules.

\subsubsection{Rewrite rules} A simple (unconditional) rewrite rule contains information on how to rewrite a formula (often to a simpler form).  A rewrite rule can also be \emph{conditional} where the results of the rewriting depends on the circumstances.  In \Fig{fig:rules}, Rules \texttt{isEmpty\_nil\_rewrite} and \texttt{isEmpty\_cons\_rewrite} are unconditional where Rule \texttt{isEmpty\_rewrite} is conditional.

\subsubsection{Inference rules} 
\begin{figure}[!h]
  \centering
  \includegraphics[scale=0.4]{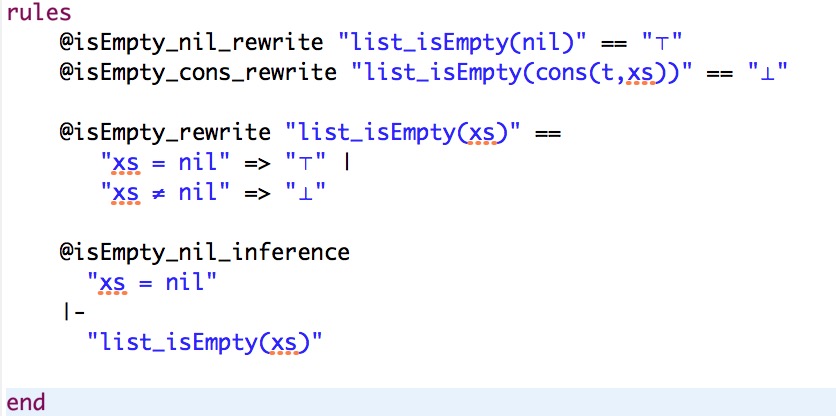}
  \caption{Example of proof rules for \texttt{List} Datatype}
  \label{fig:rules}
\end{figure}

An inference rule contains a list of \emph{given conditions} (possibly empty) and the inferred clause.  In \Fig{fig:rules}, Rule \texttt{isEmpty\_nil\_inference} is an example of inference rules. An inference rule can be applied backward or forward (by default, it is applicable in both direction).

\section{Changes to the Rodin Core}
\label{sec:changes-rodin-core}

\subsection{Compatibility of Formula Factories}
\label{sec:comp-form-fact}
The first main part of the work is to ensure that the Theory plug-in verifies the construction of the AST formula to guarantee the compatibility of the formula factories.  Since the formula factories are basically constructed from the core (standard) factory with mathematical extensions (datatypes, operators), compatibility between different formula factories is reduced to equalities between datatypes and operators. 
\begin{itemize}
\item \textbf{(Inductive) Datatype}: Two inductive datatypes are the same if they have the same signature, i.e., name, type parameters, constructors, destructors.

\item \textbf{Axiomatic Datatype}: Two axiomatic datatypes are the same if they have the same signature, i.e., name.

\item \textbf{Operators (axiomatically defined or with direct/inductive definitions)}: Two operators are the same if they have the same signature, i.e., name, and arguments (including argument types).  Note that this comparison does NOT take into account the actual definitions/properties of the operators.
\end{itemize}
With the above definition, the Theory plug-in correctly compares the formula factory in constructing proofs and checking reusability of existing proofs.

\subsection{Supporting Infix Predicate Operator}
\label{sec:supp-infix-binary}
This particular additional functionality is mainly for improving the usability of the Theory plug-in.  We have implemented support for introducing \emph{Infix predicate operator}.  For example, consider the operator $smr$ (smaller than) for a datatype $\emph{Real}$ numbers.  With prefix operator, for expressing $x$ smaller than $y$, we write \[smr(x, y)\] which is unnatural.  With the support for infix operator, we can write the formula as \[x \wide{smr} y\] which improves the readability of the formal text.  This support requires modification of both the Rodin platform core and the Theory plug-in core.

\subsection{Supporting Type Specialisation}
\label{sec:supp-type-spec}

The Theory plug-in has to instantiate several generic datatypes and operators.
Such an instantiation means that both types and variables have to be
replaced at the same time.  For instance, the classical list operator is
defined as $\mathit{List}(T)$, but can be used as $\mathit{List}(1\upto 3)$.

The same issue also needs to be addressed by other plug-ins, such as
generic instantiation plug-ins that allow to make a generic development and
later instantiate it into another development.

Therefore, all the machinery for instantiating at the same time types and
variables has been developed withing the Rodin core in Rodin 3.x under the name
of specialisation (the \emph{instantiation} name being already used for another
purpose).  This specialisation mechanism can be applied to any type, type
environment or formula and guarantees type safety (if the input is well-typed,
then the output is guaranteed to also be well-typed).

\section{Changes to the Theory Plug-in}
\label{sec:changes-theory-plug}

\subsection{Improvement on Pattern Matching}
\label{sec:impr-patt-match}
The pattern matching facility of the Theory plug-in has been upgraded to use directly the support from the Event-B core for specialising (instantiating) formulae.  This ensures that the information for specialising formulae is type-consistent.  The following examples illustrate some consistent specialisations (which are unsupported in the previous version).
\begin{Bcode}
  $
  \begin{array}{ccc}
    Patterns & & Formulae \\
    S  & \Wide{\longrightarrow} & \pow(S) \\
    S  & \Wide{\longrightarrow} & S \cprod T \\
  \end{array}
  $
\end{Bcode}

Another important improvement on pattern matching is the implementation for matching associative operators (which is only implemented for some special cases before).  The matching for associative formulae is done based on a simple greedy algorithm.  The following examples illustrate the result of pattern matching for an associative operator, namely forward composition $;$.
\begin{center}
  \begin{tabular}{lll}
    \hline
    Patterns & Formulae & Result \\
    \hline
    $f;\{x \mapsto c\}$ & $g;h;\{y \mapsto c\}$ & $f \leftarrow g;h$ \\
                                    & & $x \leftarrow y$\\
                                    & & $c \leftarrow c$ \\
    \hline
    $e;f$ & $g;h;\{y \mapsto c\}$ & $e \leftarrow g$ \\
             & & $f \leftarrow h;\{y \mapsto c\}$
  \end{tabular}
\end{center}
Note that in the second example, another possible result for pattern matching would be $[e \leftarrow g;h, f \leftarrow \{y \mapsto c\}]$. Our algorithm only returns a single matching result that it found first in the case where there more than one possible matching.

\subsection{Supporting Unicode Typesetting for Real Number Operators}
\label{sec:supp-unic-types}
Another usability improvement that we have implemented is the support for typesetting \emph{Real} number operators in Unicode.  Instead of ASCII text, we use the following Unicode symbols for the common operators that we used in the modelling.
\begin{figure}[!h]
	\centering
	\hspace{1.5cm}\includegraphics[scale=0.26]{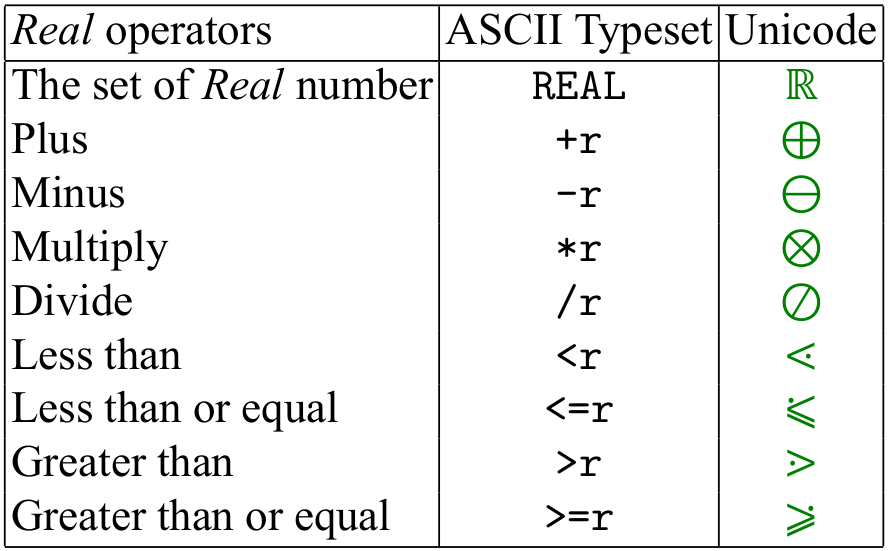}
\end{figure}

\subsection{Usability Improvement/Reimplementation Rule-based Prover}
\label{sec:impr-manu-reas}

As mentioned before, the rule-based prover of the Theory plug-in offers 5 different tactics:
\begin{enumerate}
\item Manual inferencing

\item Manual rewriting

\item Automatic inferencing

\item Automatic rewriting

\item Automatic definition expanding
\end{enumerate}

Each tactic is a wrapper around one or more reasoners.  Roughly speaking, each reasoner manages an application of a single inference/rewrite rule.  Table~\ref{tab:tactic-reasoner} shows the relationship between the tactics and the corresponding reasoner (for the pervious, i.e., \emph{v3.0}, and the current implementation, i.e., \emph{v4.0}).%
\begin{table}[!htbp]
  \centering
  \begin{tabular}{|l|l|l|}
    \hline
    Tactic & v3.0 & v4.0 \\
    \hline
    Manual inferencing & Single reasoner & Single reasoner \\
    Manual rewriting & Single reasoner & Single reasoner \\
    Automatic inferencing & Single reasoner & \emph{Multiple reasoners} \\
    Automatic rewriting & Single reasoner & \emph{Multiple reasoners} \\
    Automatic definition expanding & Single reasoner & \emph{Multiple reasoners} \\
    \hline
  \end{tabular}
  \caption{Tactics and Reasoners}
  \label{tab:tactic-reasoner}
\end{table}
The decision to implement the automatic tactics combining multiple reasoners is for \emph{usability}.  Instead bundling the effect of apply several rules into a single proof node, we separately apply the rules one-by-one.  The result is as follows
\begin{itemize}
\item The proofs are easy to understand.
\item The proofs are less prone to changes (e.g., changes in the model).
\item The proofs are easy to adapt, e.g., one can keep a part of the proofs produced automatically (by pruning) and proceed further manually.
\end{itemize}
 
\subsubsection{Reasoner Input}
\label{sec:reasoner-input}
In the previous version, the context of the proof obligation is passed as an input to the reasoner.  In this version, the context is retrieved from the origin of the proof obligation. This ensure that the same context, e.g., the formula factory, is used for the obligation and the proof.  Further more, due to the changes to the automatic reasoners described previously, the reasoner inputs for these reasoners are adapted accordingly. Table~\ref{tab:reasoner-input} summarises the differences between v3.0 and v4.0 of the tool.
\begin{table}[!htbp]
  \centering
  \begin{tabular}{|l|p{0.3\textwidth}|p{0.3\textwidth}|}
    \hline
    Reasoner & v3.0 & v4.0 \\
    \hline
    Manual Inferencing & 
                         \begin{itemize}
                         \item PO Context
                         \item Rule Meta-data
                         \item Application hypothesis (if forward inference) or \texttt{null} (if backward inference)
                         \end{itemize} &
                         \begin{itemize}
                         \item Rule Meta-data
                         \item Application hypothesis (if forward inference) or \texttt{null} (if backward inference)
                         \end{itemize}
    \\
    \hline
    Manual Rewrite & 
                         \begin{itemize}
                         \item PO Context
                         \item Rule Meta-data
                         \item Application hypothesis (if rewriting a hypothesis) or \texttt{null} (if rewriting the goal)
                         \item Rewrite position
                         \end{itemize} &
                         \begin{itemize}
                         \item Rule Meta-data
                         \item Application hypothesis (if rewriting a hypothesis) or \texttt{null} (if rewriting the goal)
                         \item Rewrite position
                         \end{itemize} \\
    \hline
    Automatic Inferencing & 
                         \begin{itemize}
                         \item PO Context
                         \end{itemize}
                    &
                         \begin{itemize}
                         \item Rule Meta-data
                         \item Forward or backward
                         \end{itemize} \\
    \hline
    Automatic Rewriting &
                         \begin{itemize}
                         \item PO Context
                         \end{itemize} &
                         \begin{itemize}
                         \item Rule Meta-data
                         \end{itemize} \\
    \hline
  \end{tabular}
  \caption{Reasoner Input}
  \label{tab:reasoner-input}
\end{table}

\subsubsection{Well-definedness}
\label{sec:well-definedness}
When instantiating a proof rule, each instantiation expression required to be \emph{well-defined}, to ensure that the resulting sequents are well-defined. In \emph{v3.0}, the well-definedness sub-goals are generated one for each instantiation expression and are mixed with the other sub-goals, and in some cases are not generated at all. We have consolidate the generation of the WD sub-goals (in \emph{v4.0}) by combing all these sub-goals into a single sub-goal (using conjunction) and add this new WD sub-goal as (always) the first sub-goal when applying a proof rule.  As a result, the proof are much less prone to changes.

\section{Future Work}
\label{sec:future-work}

In this paper, we have highlighted the major changes to the Theory plug-in and the Rodin Core, 
focusing on bringing the Theory plug-in to the latest version of the Rodin platform. In particular, the update requires some update to the core of the Rodin Platform itself, hence will be publicly available after the next release (\emph{v3.3}) of the Rodin Platform. At the same time, we also improved the usability of Theory plug-in, especially focusing on features that required changes from the Rodin plug-in core.

In the next release, we will focus our attention to the usability of the plug-in, by gathering feedbacks from its users. Some of the ideas for improvement are:
\begin{itemize}
\item Matching facility for \emph{associative and commutative operators} (currently ignoring commutativity).

\item Support for user-defined tactics 

\item Support for predicate variables in theories

\item Theory instantiation (different abstraction-level of theories).
\end{itemize}

\section*{Acknowledgements}
\label{sec:acknowledgements}
Laurent Voisin and Nicolas Beauger have been partially funded by the French Research Agency under grant ANR-13-INSE-0001 (IMPEX project).  Thai Son Hoang, Michael Butler and Toby Wilkinson are supported by the ASUR Programme project 1014 C6 PH1 104.

\paragraph{Disclaimer.} This document is an overview of MOD sponsored research and is released to inform projects that include safety- critical or safety-related software. The information contained in this document should not be interpreted as representing the views of the MOD, nor should it be assumed that it reflects any current or future MOD policy. The information cannot supersede any statutory or contractual requirements or liabilities and is offered without prejudice or commitment. 

\bibliographystyle{plain}
\bibliography{theory}

\end{document}